# Multiplierless Modules for Forward and Backward Integer Wavelet Transform

Vasil Kolev

***Abstract:*** *This article is about new architecture of a integer DWT with reprogrammable logic. It is based on second generation of wavelets with a reduced of number of operations. A new basic structure for parallel architecture and modules to forward and backward integer discrete wavelet transform is proposed.*

## Introduction

The wavelet analysis is useful for the problems in many areas of applications. This is a new systematic way for analyzing functions with a rich basis [1, 6]. In general this means that these are functions with a good time-frequency distribution and zero mean. It is also possible to miss a prototype as is for the second generation wavelets.

The discrete wavelet transform (DWT) is used in image and sound processing, noise reduction in signal, tomographic, modems digital holographic etc. Often the processors for this transformation is based on serial input and like data are input serial from the first to the last sample. These processors are effective for one-dimensional signal, but it is also possible to use them for two-and three-dimensional signals. It is possible to represent the two dimensional pictures as a sequence of one-dimensional signal with the pixels arranged with a predefined manner. The wavelet transform with lifting scheme (LS) working "in place" is described in [5, 8]. There is realization with integer arithmetic for VLSI optimising structures for JPEG2000 standard [10].
The goal of this paper is to build modules with new basic structures for integer discrete wavelet transform (integer DWT) by LS for real time processing and hardware parameters studying.

## The algorithm overview

The basic principles of LS (for analysis filter - $H(z)$ and syntesis filters - $G(z)$) are the polyphase representation:

$$H(z) = H_e(z^2) + z^{-1}H_o(z^2) \qquad (1)$$
$$G(z) = G_e(z^2) + z^{-1}G_o(z^2),$$

where the polyphase components $H_e(z^2), G_e(z^2)$ and $H_o(z^2), G_o(z^2)$ are Lauren's polynomials from commutative ring. The division with redundancy here is possible but not an unique solution. These polynomials represent the even and odd filter coefficient and form polyphase matrix [2]:

$$P(z) = \begin{bmatrix} H_e(z) & G_e(z) \\ H_o(z) & G_o(z) \end{bmatrix} \qquad (2)$$

$P(z)$ decomposition is a new form of representation with triangle matrixes, which defined the phases of prediction (**P**) and updating (**U**) (Fig.1)

### Analysis

Let the signal $S_j$ be a given signal of j-level. The requirement for decorrelation defines the problem for signal representation with minimum coefficient number [4].

- **Spit**

After using lazy wavelet the input samples are separated to even and odd areas of coefficient [8](Fig.1).

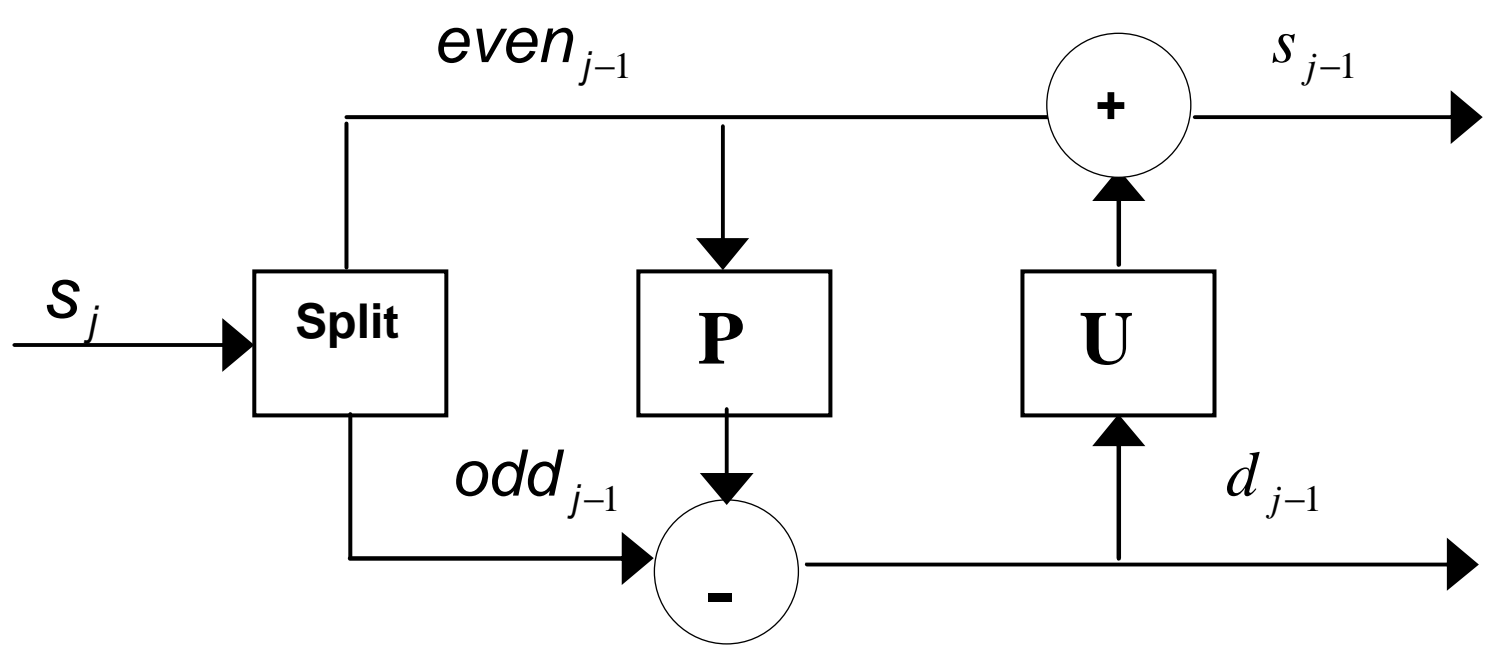

fig.1 Lifting sheme, forward transform

The half of the samples of original signal forms each one of these two areas.

$$(even_{j-1}, odd_{j-1}) := \mathbf{Split}(s_j) \tag{3}$$

- **Prediction** (**P**)

Let the signal be strongly correlated. It is possible to use an even sequence for prediction of odd sequence. The detail part $d_{j-1}$ is the difference between odd and prediction values:

$$d_{j-1} = odd_{j-1} - \mathbf{P}(even_{j-1}) \tag{4}$$

It is possible to use triple samples $s[2n], s[2n+1], s[2n+2]$ to line interpolation with a prediction phase (**P**) equal to average value of even samples. The difference between odd and predicted value forms the wavelet. It is utilized an operator for rounding (floor function $\lfloor x \rfloor$- is the greatest integer less than or equal to x*)*. It is found the following wavelet equation:

$$d_j[n] = s_{j-1}[2n+1] - \left\lfloor \frac{s_{j-1}[2n] + s_{j-1}[2n+2]}{2} \right\rfloor \tag{5}$$

If the odd value coincide with predicted value, then wavelet coefficient is zero .

- **Update** (**U**)

The main goal of phase updating (**U**) is to find from calculated wavelet coefficients the scalable function for saving the current and average of the wavelet for all level of decomposition. The coefficient $s_{0,0}$ is average value of signal for last level. The updating represents the scaling function for the next lifting step.

$$s_{j-1} = even_{j-1} + \mathbf{U}(d_{j-1}) \tag{6}$$

From the above condition the scaling coefficients can be represented by the expression:

$$s_j[n] = s_{j-1}[2n] + \left\lfloor \frac{d_{j-1}[n] + d_{j-1}[n-1]}{4} \right\rfloor \tag{7}$$

***Reconstruction***

From (6) the inversion can be found of the step of updating, which result is an even value from the restored sequence of a given level:

$$even_{j-1} = s_{j-1} - \mathbf{U}(d_{j-1}) \tag{8}$$

From $s_{j-1}$ and $d_{j-1}$ can just be found the inversion of prediction step (4), which result is an odd value from restored sequence of given level:

$$odd_{j-1} = d_{j-1} + \mathbf{P}(even_{j-1}) \tag{9}$$

The reconstruction signal $\tilde{s}_j$ is described as a merge of even and odd value:

$$\tilde{s}_j = \mathbf{Merge}(even_{j-1}, odd_{j-1}) \tag{10}$$

**The new basic structure**

The algorithm of realization of LS requires to store in the same time some values. The proposed in [5] basic architecture consists from two adder, three registers and one multiplayer. Such configuration requires more resource and is inconvenient for updating. In [10] is propose a sructure with one multiplayer, one adder, register and multiplexor. The disadvantages are the requirements of a big number of registers for making multiplying, with which the time of arithmetic operations increase and the possibility the samples be processed only in a serial way.

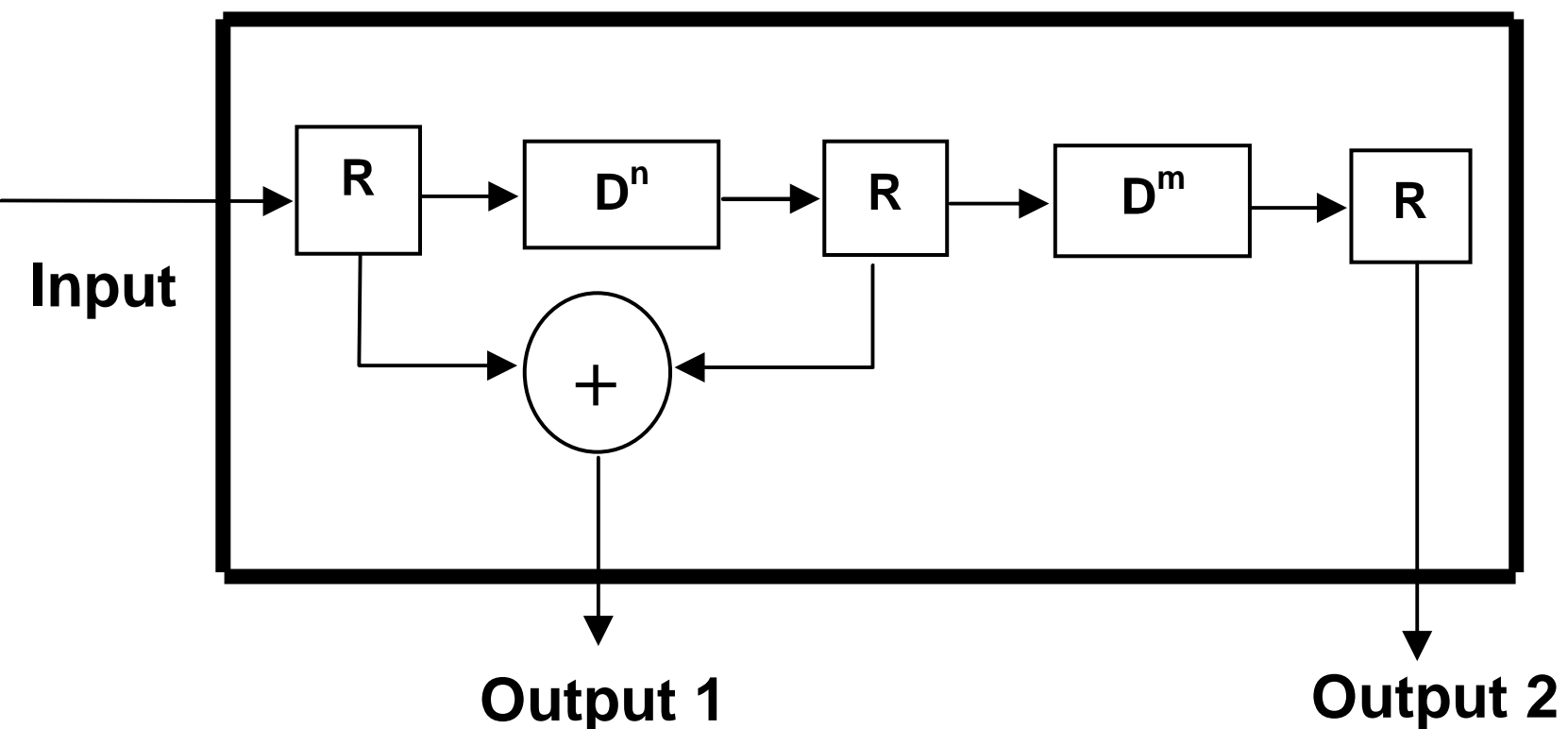

Fig.2 The new processing element

The new structure (Fig.2) is consisted of two programmable delays ($D^m$ и $D^n$), tree registers(R) and one adder. This structure is in the basis of builder integer DWT.

**Modules describing**

**1. Modul analysis** (Fig.3).

The input signal is 8-bits. The sum of even values ($s_{J-1}[2n]$ and $s_{J-1}[2n+2]$) is stored in register making division by two as a shift one bit right. The result from ($s_{J-1}[2n+1]$) is substracted with a pass over to two-complement code. At the output is obtained the wavelet values $d_{J-1}[n-1]$ and $d_{J-1}[n]$ which are summated. If the sum is negative, the operation division by four is combined with one bit correction and the result is added with even value $s_{J-1}[2n]$. That gives the scaling values $s_{J-1}[n]$. The blocks-dashed line is used in the new basic structure.

**2. Modul reconstruction** (Fig.4)

There are wavelets $d_{J-1}[n]$ and scaling function $s_{J-1}[n]$ samples at the input. The block enclosed with dashed line is made with new bases structure. If the sum of values $d_{J-1}[n]$ and $d_{J-1}[n-1]$ is negative, it is necessary to make correction of the result with one bit for error eliminating from next operation.The register with the sum is shifted two bits to the right for division of four.

Next is the difference between the result of the shifting and $s_{J-1}[2n]$, and that give even values s[2n] stored in an additional register. These two values are summarized and it is made a division by two for this result with a shifting of register one bit to the right. The output of division is summed with $d_{J-1}[n]$ and that gives the odd value s[2n]. By using a multiplex it is possible to have reconstruction values ( $\tilde{s}_{J-1}$).

**The Hardware implementation of the modules**

The reprogrammable logic changes real time the computer system with reconfiguration of system level and with change of corresponding data and instruction for the individual processing element (PE). This logic works with greatly frequencies than that for the digital signal processor (DSP). The disadvantage of this logic is the limit on number of logical operations. It is possible to use the advantages of hardware description languages and to apply the method of design from down to top level. It is available the error correction and the possibility to make changes in whole chip structure. Here it is used the high-level hardware description language VHDL for design description. Xilinx Foundation is used for place and routing. The simulations are made with ModelSim5.6a of MentorGraphics. The test is with consideration of the application of the modules, which will work with integer positive samples. A signal with 64 samples (Fig.5) is generated whose values have with a normal distribution.

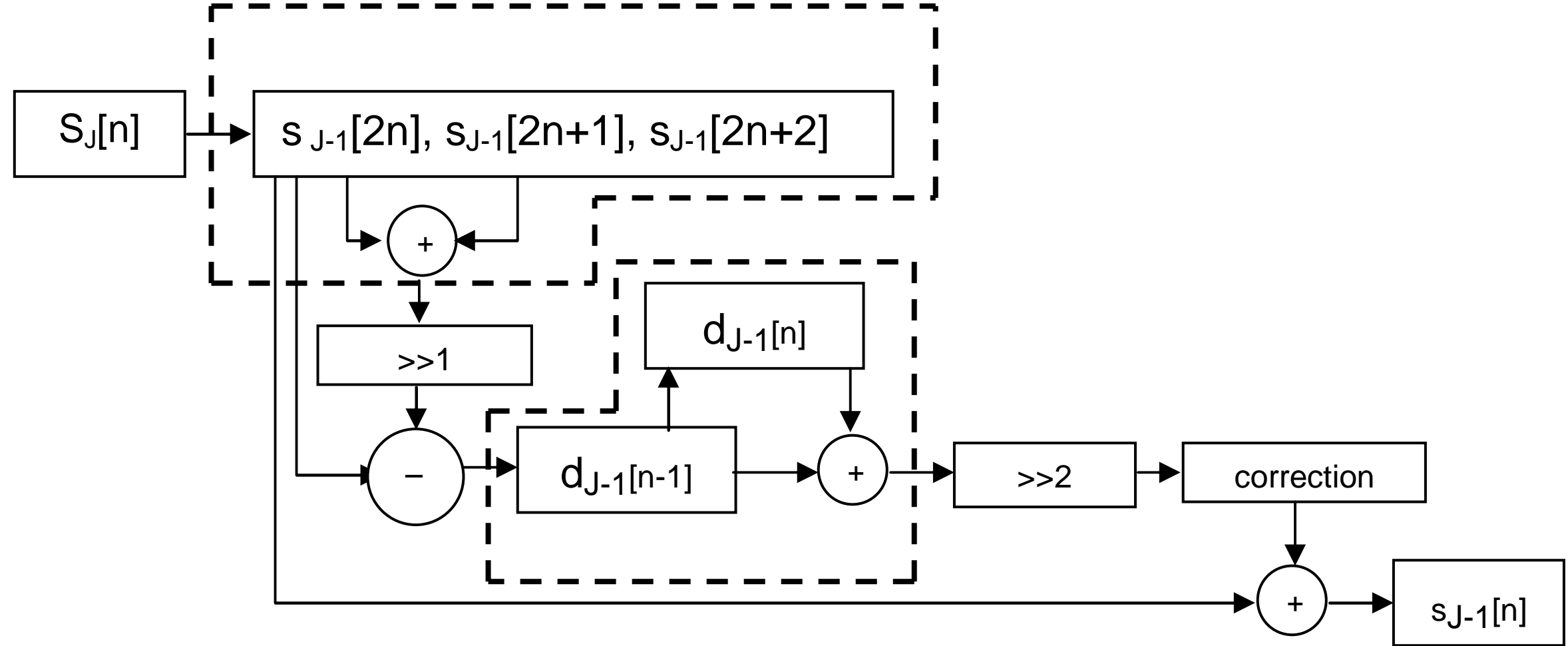

Fig. 3 Architectures of forward integer DWT

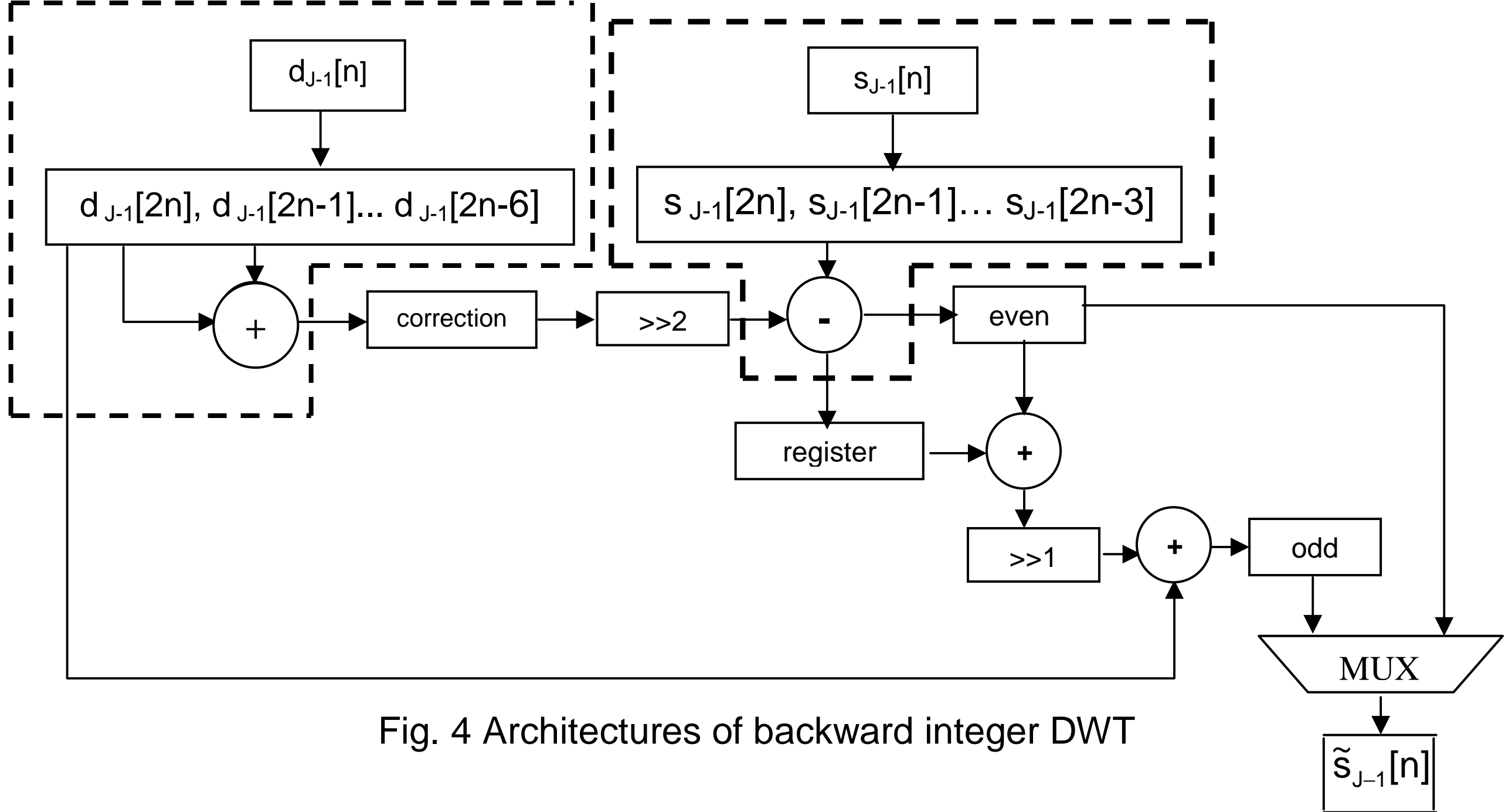

Fig. 4 Architectures of backward integer DWT

After the optimization the designed modules have the following characteristic (Table 1). The designed with the proposed new basic structure modules gives the possibility of decreasing the hardware requirements for generation of the outputs of forward ($s_{J-1}[n]$, $d_{J-1}[n]$) and backward ($\tilde{s}_{j-1}$) integer DWT (Табл.2).

**Table 1**

| **Module Analysis**<br>**Target Device**: Virtex Excv200e-pq240-8 | **Module Reconstruction**<br>**Target Device:** Spartan2 xc2s150-fg256-6 |
| --- | --- |
| #Work frequency (MHz) :100<br># Registers : 30<br>8-bit register : 30<br># Adders/Subtractors : 5<br>8-bit subtractor : 1<br>8-bit adder : 4 | #Work frequency (MHz) :100<br># Registers : 21<br>9-bit register : 21<br># Adders/Subtractors : 6<br>9-bit adder : 6 |

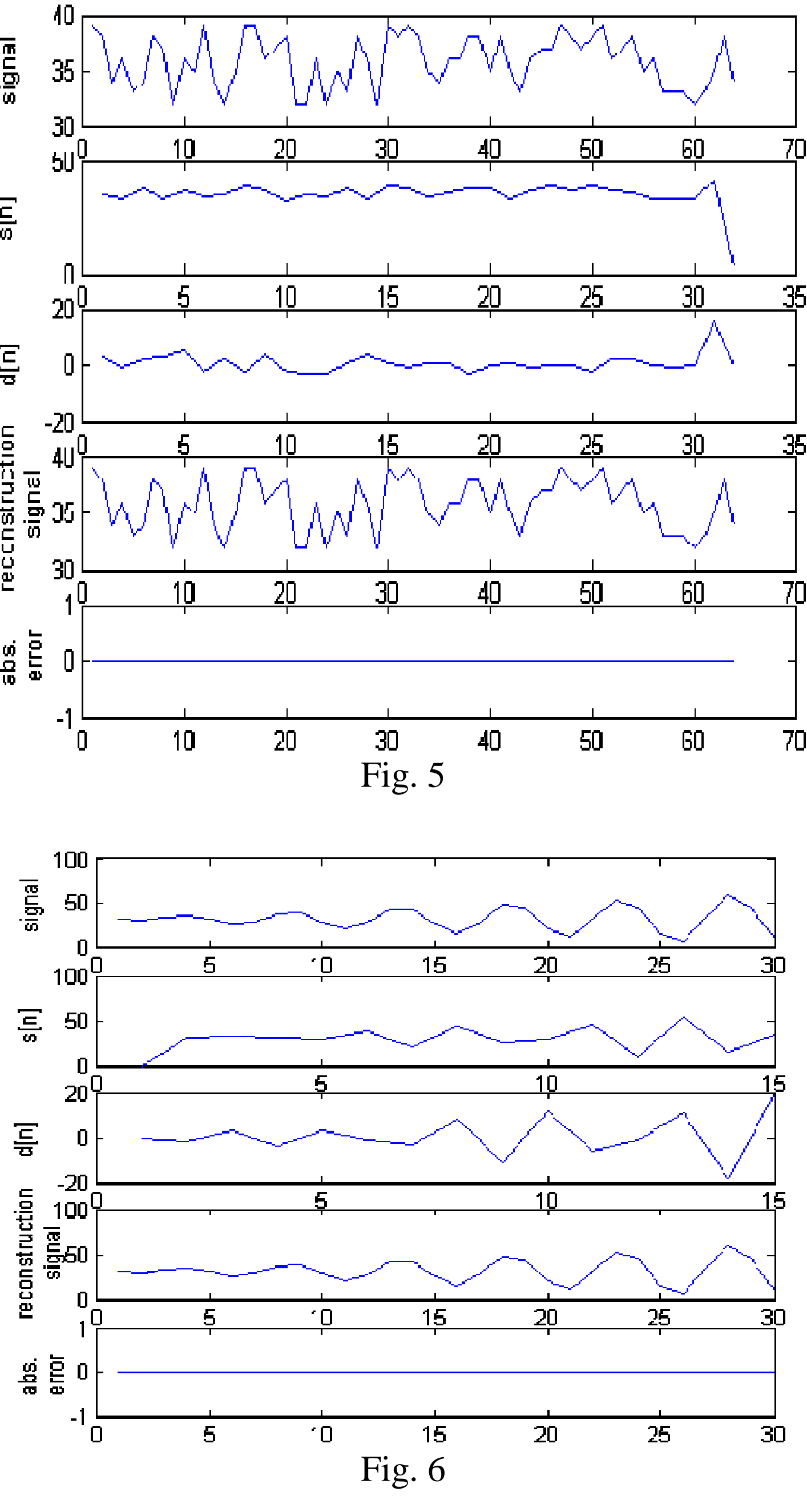

Fig. 5

Fig. 6

**CONCLUSIONS AND FUTURE WORK**

It is possible to make the following conclusions from the designed modules and from the achieved result:

- Parallel processing with the proposed block structures simplified the hardware implementation for the arithmetic operation realization (Табл.2);

**Table 2**

| Filter (5/3) | This work | Kishore A. [5] |
|---|---|---|
| Adders | 4 | 8 |
| Shifters | 2 | 4 |

**Table 3**

| This work | DSP [9] Fl. point | Dwt[10] Fpga |
|---|---|---|
| 12µs | 400µs | 20µs |

- The forward and backward integer DWT by Lossless scheme have the same calculation complexity;
- The possibility for parallel and serial operations with data in a clock period by using the state chart;
- It is possible to be used processing of data that have not only the power two;
- The standard methods for (5, 3) filter bank require 8 operations while the LS only 5. This make useful and effective the integer values filters for calculating speed increase;
- A given integer value of signal after transformation is lossless transformation (Fig.5);
- The architectures for fixed point arithmetic (the proposed modules and these in [10]) with comparison of these with floating point arithmetic gives some higher speed for processing of line with 256 samples with 8–bit accuracy (Table 3);

In the future work it is expected to be applied the modules for IDWT with several level and analyzing with LS algorithm for sound and image compression. This is a possibility for modules ASIC implementation.

**ABOUT THE AUTHOR**

Vasil Kolev, Institute of Information and Communications Technologies, Bulgarian Academy of Sciences, E-mail: kolev_acad@abv.bg.